# Effects of Oxidation on the Tribological Properties of Diamond Sliding Against Silica. Insights from *Ab initio* Molecular Dynamics

*Huong T. T. Ta[1], Nam V. Tran[1], and M. C. Righi[1]\**

[1]Department of Physics and Astronomy, University of Bologna, 40127 Bologna, Italy

Corresponding Author:

M.C. Righi – E-Mail: clelia.righi@unibo.it, Department of Physics and Astronomy, University of Bologna, 40127 Bologna, Italy.

**Abstract**

Tribological phenomena such as adhesion, friction, and wear can undermine the functionality of devices and applications based on the diamond-silica interface. Controlling these phenomena is highly desirable, but difficult since extrinsic factors, such as the surface termination by adsorbed species, can deeply affect the reactivity of diamond and its resistance to wear. In this work, we investigate the effects of diamond oxidation by massive *ab initio* molecular dynamics simulations of silica sliding against diamond surfaces considering different surface orientations, O-coverages, and tribological conditions. Our findings reveal a dual role of oxygen that depends on coverage. At full coverage, the adsorbed oxygen is very effective in friction and wear reduction because the repulsion with the silica counter-surface prevents the formation of chemical bonds across the interface. At reduced coverage and high pressure, Si-O-C bonds are anyway established. In this situation the presence of oxygen results detrimental as it weakens the surface C-C bonds making the surface more vulnerable to wear. Indeed we observed atomic wear on the C(110) surface at 50% O-coverage under harsh tribological conditions. The mechanisms of friction reduction and atomistic wear are explained through the analysis of the electronic properties and surface-surface interactions. Overall, our accurate *in silico* experiments shed light into the effects of adsorbed oxygen on the tribological behavior of diamond and show how oxidized diamond can be worn by silica.

Keywords: Friction, diamond wear, atomistic mechanisms, *ab initio* Molecular Dynamics

## 1. Introduction

Due to its outstanding stability, hardness, ultra-low coefficient of friction and wear rates,[1–3] diamond is a material of choice for many applications including coatings, bearings, cutting tools and micro-electromechanical systems. Understanding the friction and wear mechanisms of diamond is, thus, highly relevant to optimize its potential in many applications and devices. However, the tribological properties of diamond are deeply affected by extrinsic factors such as the surface termination upon interaction with adsorbed species during chemical vapor deposition,[4] etching processes,[5] or exposure to atmospheric gases.[6–8] In particular, oxygenation of diamond surfaces has been considered as an influential technique for polishing and etching processes.[9] The presence of oxygen can, in fact, modify the diamond surface chemistry and its electronic structure,[9] altering its reactivity,[10] electron affinity,[11] and key electrical/photonic properties.[12,13] In addition, the role of oxygen in reducing friction and wear of diamond has been experimentally observed,[14–17] while in harsh conditions, the presence of oxygen on diamond surfaces seems to make them easier to be worn.[18] However, a comprehensive understanding of the atomic mechanisms of friction and wear of the oxidized diamond surface is still lacking, despite its importance for the quality improvement in diamond applications.

Different mechanisms to explain the ultra-low friction and wear of diamond have been reported.[19–23] One key mechanism is the passivation of reactive dangling bonds, which can reduce adhesive friction in diamond sliding contacts.[14,22,24,25] The passivating species are usually H and OH fragments from the dissociative adsorption of $H_2O$ in humid environment.[22,26,27] Meanwhile, C-O and C=O structures can be dominant when molecular oxygen decomposes on the surface,[10,28] or OH groups are dehydrogenated in tribological conditions or at elevated temperatures.[26] The lower activation energy for oxygen dissociation than that of $H_2$ and $H_2O$, suggests that O-terminating diamond surfaces are more common than the H- and OH-terminated ones.[29] At silica-diamond interfaces, O-terminated diamond surfaces present lower adhesion and larger separation than H-passivated ones due to the large atomic size of oxygen which induces higher steric hindrance compared to hydrogen.[30] Nevertheless, the tribological performance of oxygenated diamond surfaces still heavily depends on passivation concentrations.[16]

In addition to friction, the surface rubbing at high applied loads can cause wear. Studies on diamond-like carbon (DLC) films indicated that wear can be formed through the transformation from diamond-like to graphite-like carbon,[19] which later was verified by MD simulations by a pilot atom concept.[23,31] Furthermore, wear can be initiated under different forms of carbon such as carbon chains,[32] atom-by-atom, and carbon sheets.[33] The mechanisms of wear have been widely studied on clean diamond surfaces. However, adsorbed species can deeply modify the chemical, electronic, and structural properties of diamond surfaces with consequent modifications of the tribological properties. Under oxygenation, carbon removal can proceed through the desorption of CO or $CO_2$,[8,9,34] suggesting the essential role of oxygen in the formation of wear. Interestingly, the wear of diamond surfaces was observed when sliding against softer materials such as silica or silicon.[31,35–39] The chemical adhesion and wear is initiated by the Si-C/C-O-Si chemical bonds between two mated surfaces.[31] However, a systematic study of the wear of oxygenated diamond sliding against silica considering different coverage concentrations, surface orientations, and tribological conditions is still lacking.

In this work, we performed *ab initio* molecular dynamics (AIMD) simulations of diamond sliding against silica to provide insights into the tribological properties of oxygenated diamond surfaces. We consider the three most common low-index diamond surfaces, i.e., the C(001), C(110), and the Pandey reconstructed C(111) (*R*-C(111)) surfaces, with two different oxygen coverages of 50% and 100%. The simulations were performed at 1 GPa to investigate the frictional performance of oxygen passivated diamond surfaces and compare it with hydrogen-passivated ones under the same tribological conditions.[30] AIMD simulations at a hasher conditions were also performed to activate wear, the atomistic mechanisms of which have been explained on the basis of the electronic structure analysis.

The presented study relies on fully *ab-initio* simulations, which ensure an accurate description of the bond-breaking and forming events activated under tribological conditions, large-size models (up to ~400 atoms) are adopted for a realistic description of the amorphous silica surface, long simulation runs (15 ps for each run) are performed for several systems in parallel. The computational effort spent for the production of the presented results puts our work at the frontier of what can be currently done with massive fully *ab-initio* calculations. We expect that the outcomes of our simulations will be relevant for enhancing the understanding of the processes occurring during the silica-driven polishing of diamond surfaces.

## 2. Simulation method and models

AIMD simulations were performed using a modified version of the Quantum Espresso package[40] which allows simulating the tribological conditions. The code has been successfully used for studying the tribochemistry of different systems, including diamond-silica interfaces.[30] The generalized gradient approximation (GGA) with the Perdew-Burke-Ernzerhof (PBE)[41] method was used as the exchange-correlation functional. The plane-wave basis set was used to expand the electronic wave function, and the core electrons were represented by ultrasoft pseudopotentials with the cutoff energies of 30 Ry for the wave function and 240 Ry for the charge density, respectively. The gamma point was used for the Brillouin zone sampling to compromise the computational cost considering the large models and a long simulation time of 15 ps. A semi-empirical correction by Grimme (D2)[42,43] was adopted to account for the long-range van der Waals interactions. The Verlet algorithm with a timestep of 20 a.u. (~1 fs) was used.

The amorphous silica was built as technically described in our previous work.[30] Three diamond surfaces with two different oxygen concentrations are mated against silica, for a total of six silica-diamond interfaces. The initial structures of these silica-diamond systems are reported in Figure S2 in the Supporting Information (SI). An external force corresponding to a load of 1 GPa was applied along the z-direction. After the relaxation and equilibration at a temperature of 300 K, the silica was slid against the diamond surfaces with a constant velocity of 200 m/s. Simulations at harsh conditions of 10 GPa and 600 K were also performed to represent extreme working conditions, where chemical processes such as C-C bond breaking can be activated in the simulated time interval. The movies of the AIMD simulations of silica sliding against oxidized diamond surfaces are provided as a part of the SI.

To understand the nature of the silica-diamond interaction, the adhesion between the two surfaces was calculated as a function of the surface separation. This perpendicular potential energy surface (P-PES) was estimated by calculating the adhesion energies at four different lateral positions of silica on diamond surfaces, the lowest one was considered as the representative of the P-PES, and the others were used to estimate the associated error bar. To make a comparative conclusion about the effect of oxygen and hydrogen passivation on the frictional properties of the silica-diamond systems, the P-PESs were calculated for both O- and H-passivated diamond surfaces.

## 3. Results and discussion

### 3.1 Diamond surfaces

The optimized structures of the six diamond surfaces studied in this work are presented in **Figure 1**. They include the C(001), C(110), and *R*-C111 surfaces covered with oxygen in two different concentrations of 50 and 100%. Based on the literature reported on the oxidation chemistry of diamond surfaces, oxygen terminations are formed as a result of oxygen dissociative adsorption, forming ether or ketone configurations depending on the surface orientation and the oxygen coverage.[8–10,28] Particularly, the ketone configuration dominates in the case of the C(001) surface, at both 50% and 100% coverages.[9,10] For the *R*-C(111) surface, the ether configuration is more favorable at low coverage, while at higher coverage the ketone group prevails.[8] In the C(110) surface, there is a coexistence of the ether and the ketone groups (**Figure 1**).[28] Active carbon sites are present on only C110-50%O and C001-50%. The former contains fewer active carbon sites than the latter due to the combination of the kenton and ether configurations. The surface configurations as well as the oxygen coverage govern the surface chemistry and the reactivity of diamond in dynamics triblogical simulations.

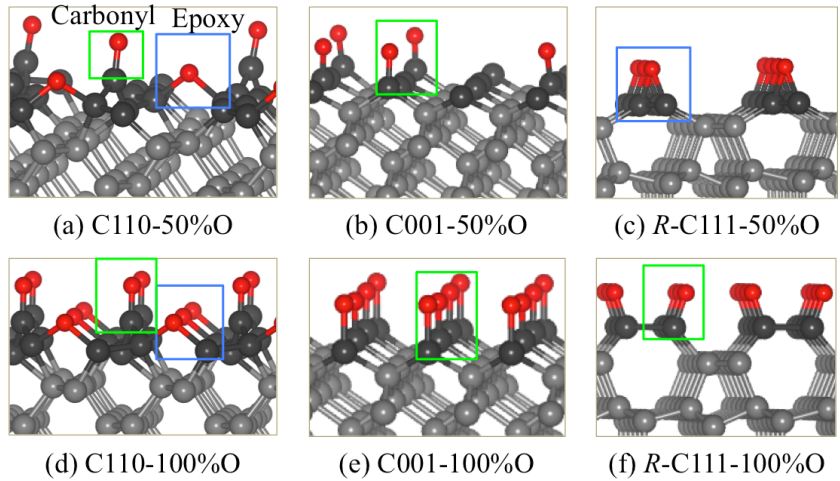

| (a) C110-50%O | (b) C001-50%O | (c) *R*-C111-50%O |
| (d) C110-100%O | (e) C001-100%O | (f) *R*-C111-100%O |

Figure 1. Optimized structures of the 50% and 100% oxygen terminated C(110), C(001), and *R*-C(111) surfaces. The darker balls show the carbon atoms at the top layer. The green and blue boxes

mark the carbonyl and epoxy configurations of the diamond surfaces. Top and side views of the corresponding diamond surfaces are shown in **Figure S1**.

## 3.2 Tribological properties of silica – oxidized diamond interfaces

Six silica-diamond systems were relaxed and equilibrated under the load of 1 GPa and the temperature of 300 K for 1 ps before being slid for 15 ps. The structures of the six simulated systems after the relaxation at 1 GPa are reported in **Figure S2,** which shows that only hydrogen bonds are formed between H of the silanol groups and O of the diamond across the silica - diamond interfaces. The snapshots of atomic structures of the silica-diamond systems during the AIMD sliding simulations are shown in **Figure 2**. During the sliding at 1 GPa, except for the C001-50% system, no chemical bonds are formed (**Figure 2a-e, Movie 1-5**). This result apparently indicates that 50% and 100% oxygen passivation can effectively hinder the chemical bond formation across the silica-diamond interface. The chemical bonds are observed only in the C001-50%O system, where OH bond dissociation followed by Si-O-C bond formation occurs, making this system the most reactive one at 1 GPa. The reactions on C001-50%O system occur in a three-steps process (**Figure 2f-i, Movie 6**): (1) Hydrogen bond formed between H1 of the silica and O2 of the carbonyl groups on the C(001) surface; (2) O1-H1 bond from the silanol group is dissociated, forming a new O2-H1 bond on the C(001) surface and leaving one non-bridging oxygen atom O1; (3) the newly formed non-bridging oxygen is chemically active, forming a C1-O1-Si bridge at the silica-C(001) interface (Figure 2g). During sliding, the relative movement of the silica results in the stretching and dissociation of the Si-O bonds. Despite the chemical interactions present in this system, the sliding leads to the breaking of Si-O bonds rather than C-O or C-C bonds, indicating that all diamond surfaces remain intact, and no wear is generated at 1 GPa in the simulated time interval.

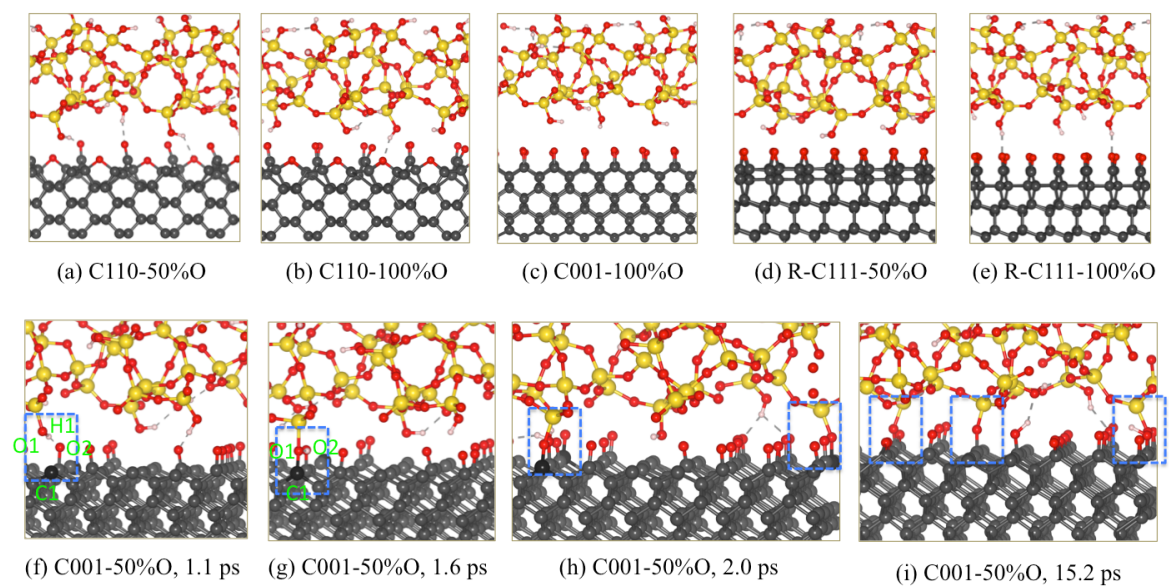

Figure 2. Snapshots of chemical events occurring at the silica-diamond interface during the tribochemical simulation at 1 GPa and 300 K.

There are two critical factors leading to the formation of C-O-Si chemical bonds at the interface: (i) the existence of the carbonyl groups, which contain chemically active non-bridging oxygen atoms that can capture H atoms from the silanol groups, and (ii) the availability of un-terminated carbon atoms that can form Si-O-C bonds immediately after the OH bond dissociation. These two factors only coexist on C001-50%O and C110-50%O surfaces. For the *R*-C111-50%O, oxygen atoms of the ether configuration terminate all active carbon sites of the topmost layer. Thus in this case, no C atoms remain un-passivated and the C-O-C configuration is less chemically active than the C=O one (**Figure 1e**). In the case of C110-50%O, the combination of carbonyl and epoxy configurations involves more carbon atoms at the topmost layer than the C(001) surface, making the C110-50%O surface less active towards silica than C001-50%O. This also explains why there are no chemical bonds formed at the silica-diamond interfaces in 100%O coverage systems, as no active carbon atoms are available on the surfaces. The results clearly indicate that the full oxygen coverage of the diamond surface effectively prevents the chemical bond formation across the interfaces.

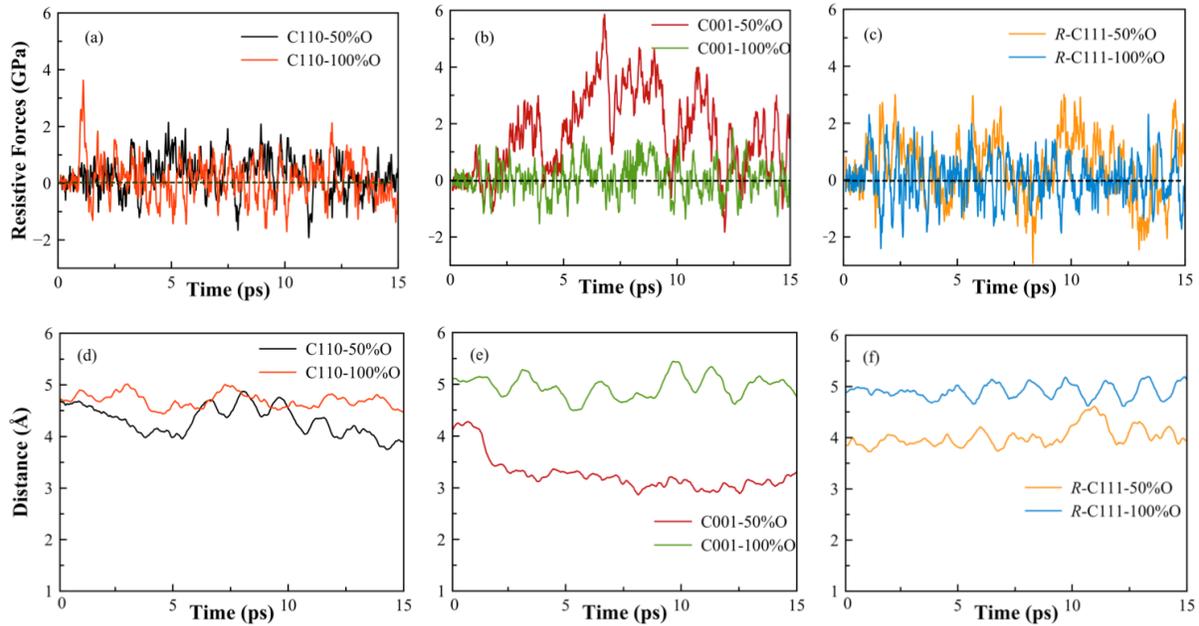

Figure 3. Time evolution of resistive forces per area (in GPa) and interfacial separation (in Å) recorded during the tribological simulations of silica-diamond interfaces.

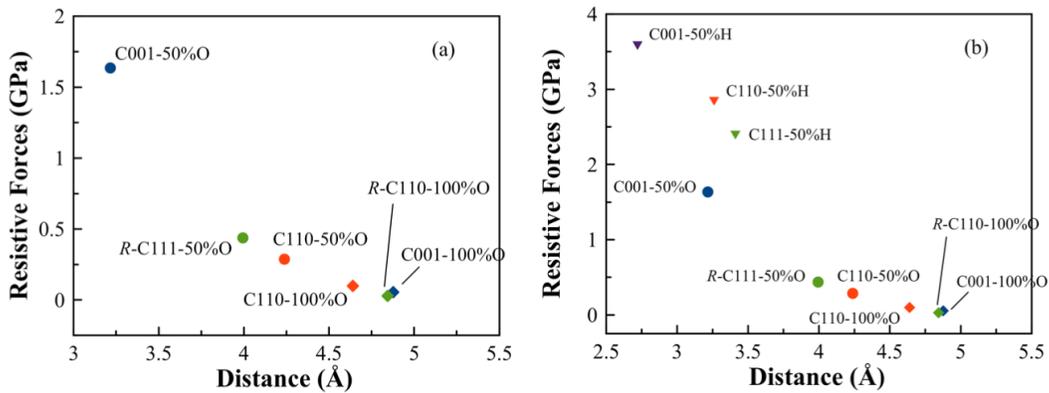

Figure 4. Mean resistant forces per area (in GPa) and interfacial distance of the silica sliding against O-terminated diamond (a) and the comparison of the mean resistive forces and interfacial

distances in O-terminated and H-terminated diamond surfaces. Color assignment: C(001) (Blue), C(110) (red), and C(111) (Green). The data of the C001-50%H, C110-50%H, C111-50%/100%H in Figure 4b is collected from Cutini et al.[30]

The calculations of resistive stress and interfacial distances recorded during the simulation are presented in **Figure 3,** while the average values over the sliding period of 15 ps are reported in **Figure 4**. It clearly shows that clear gaps from 4 to 5 Å are maintained in five of the six simulation systems, except for the case of C001-50%O. There is a correlation between the resistive force and the interfacial distance, i.e., the larger the interfacial gap, the lower the resistive force. Accordingly, the interfacial distances (~5 Å) with the lowest resistive forces are obtained for all the 100%O coverage systems, suggesting that 100%O coverage is the best value for adhesion reduction. This is in good agreement with the literature which revealed that a passivation concentration greater than 50% is needed to secure friction reduction.[16] The full passivation of the surface is effective in preventing chemical bonds across the silica-diamond interfaces, thus reducing adhesion and resistive forces. On the other hand, the highest resistant forces with the interfacial distance reduced to ~3.0 Å (**Figure 3e**) were experienced in the C001-50%O system. In this case, the Si-O-C chemical bond formation, as shown in **Figure 2f-i**, plays a key role in increasing adhesion and resistant forces.[44] It is worth mentioning that the separation distance is calculated between the lowest Si atoms and the topmost carbon layer. Thus, a separation distance of at least ~3.12 Å, which is the sum of the Si-O and C-O bond lengths, is necessary to establish chemical bonds across the interface.

Compared with the H passivation, the full oxygenation of the diamond surfaces provides comparable friction reduction to the fully hydrogenated systems. Meanwhile, for the half coverage, oxygenation can provide even better friction reduction than hydrogenation as lower resistive stress and higher interfacial distances are obtained. The larger gaps in O-terminated diamond can be due to the larger size of oxygen, resulting in a greater hindrance effect compared to that of hydrogen. Another reason is that, O-terminated diamond and silica surfaces contain oxygen atoms at the interface which promote electrostatic repulsion between the two surfaces, thus keeping the two surfaces apart. As shown by the Bader charges reported in **Table 1**, oxygen atoms carry negative charges (the average values from -0.65 – 0.95$e$ in diamond surfaces and -1.53$e$ in silica) that create repulsive interaction between the two opposed surfaces. Whereas, in H-terminated systems, the attraction between oxygen-rich silica and hydrogen atoms on the diamond surface keeps the two surfaces at close distances. Therefore, full oxygenation of the diamond surface can provide even lower resistive forces than the corresponding hydrogenation.

Table 1. Average Bader charges ($e$) of atoms in diamond surfaces and silica. The averages are calculated considering the same atomic types in the systems.

| System | C(110) | | | C(001) | | | $R$-C(111) | | | Silica |
|---|---|---|---|---|---|---|---|---|---|---|
| | 50%O | 100%O | 50%H | 50%O | 100%O | 50%H | 50%O | 100%O | 50%H | |
| C/Si* | 0.07 | 0.11 | 0.03 | 0.08 | 0.11 | 0.03 | 0.07 | 0.14 | 0.03 | +3.16 |
| C(=O)/C(-H) | +1.02 | +1.01 | -0.03 | +0.89 | +0.71 | | | +0.91 | -0.02 | |
| C(-O-C) | +0.39 | +0.37 | | | | | +0.35 | | | |
| O(=C)/O(-Si)* | -0.95 | -0.91 | | -0.87 | -0.65 | | | -0.90 | | -1.53 |
| O(-C) | -0.80 | -0.78 | | | | | -0.72 | | | |
| H(-C)/H(-O)* | | | +0.06 | | | +0.06 | | | +0.05 | +0.67 |

* only in silica

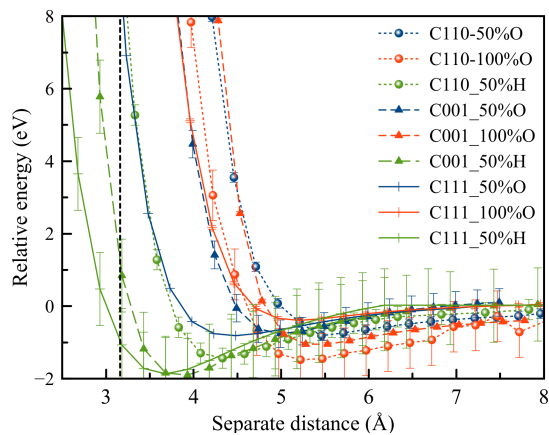

Figure 5. P-PESs of the silica-diamond systems as a function of separation distance. The separation distance is measured from the average Z-coordinates of the lowest Si atoms to the highest C atoms.

To provide more insights into the nature of silica-diamond interactions in O- and H-terminated systems, we calculated the P-PESs of the six systems studied in this work together with three H-terminated systems from our previous work.[30] As depicted in **Figure 5**, the hydrogen-terminated systems (green curves) present deeper minima at the shorter separation than all oxygen-terminated systems. This can be due to the charge distribution that promotes the hydrogen interaction between the H-passivated diamond and the O-rich silica and attracts the two surfaces at a closer distance. Meanwhile, the minima of oxygen-terminated systems (blue and red curves) shift to higher separate distances, suggesting that the repulsion is dominated by the oxygen layer. Higher repulsion in oxygen-terminated systems at short distances keeps the silica and the diamond surfaces at a certain separation.

### 3.3 Initial stage of diamond wear in harsh conditions

Hash conditions of 10 GPa load and 600 K are applied to facilitate the formation of wear in silica-diamond systems within the simulation time interval. The optimized structures at 10 GPa are shown in **Figure S3.** Only two systems, C110-50%O and C001-50%O, among the six systems, show chemical bonds across the interface. In addition, literature has reported that the C(111) surface is the hardest facet of diamond to be polished.[45] Therefore, the AIMD simulations in the harsh conditions are carried out for the C(110) and C(001) systems.

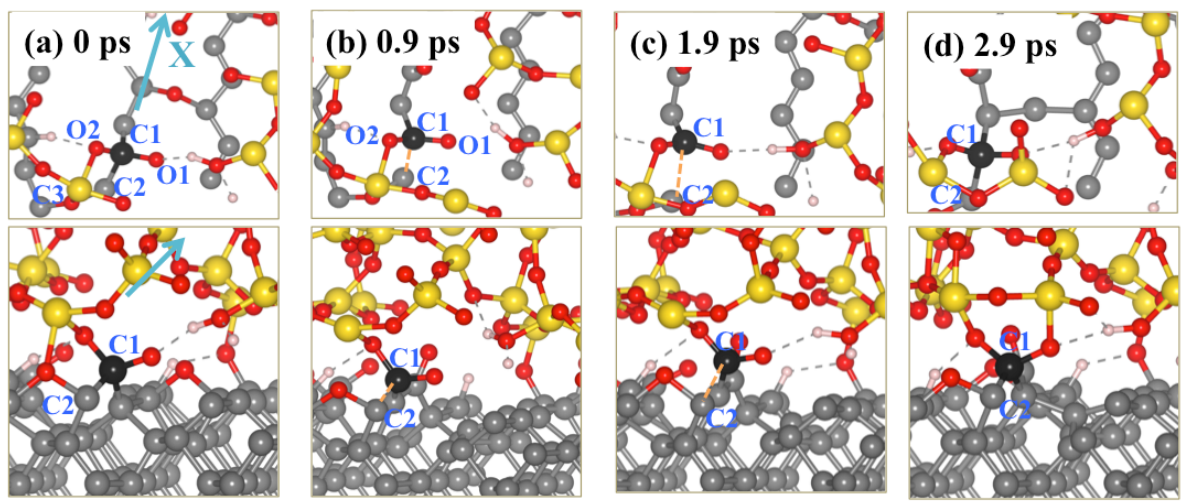

Figure 6. Top (first row) and side (second row) views of the C-C bond dissociation in the silica-C110-50%O system under the harsh conditions. The bold ball represents the detached carbon atom. Only one topmost layer of the C(110) surface and one bottommost layer of the silica slab are shown in the top views.

As can be seen in **Figure 6**, in the C110-50%O system, after the relaxation at 10 GPa, the bonding of C1 and C2 to oxygen atoms makes the C1-C2 bond weaker with a distance of 1.67 Å (Figure 6a). This distance is larger than the average bond length of 1.35-1.54 Å for other C-C bonds on the top layer.[31] The C-C bond breaking occurs at 0.95 ps of the equilibrium process. The C1-C2 distance reaches a value of 2.51 Å after sliding for ~0.9 ps (**Figure 6c**), indicating the complete dissociation of the C1-C2 bond. However, under the applied load, C1 and C2 recombine at ~ 1.9 ps and remain bonded. The further sliding only results in the dissociation of Si-O/C-O bonds.

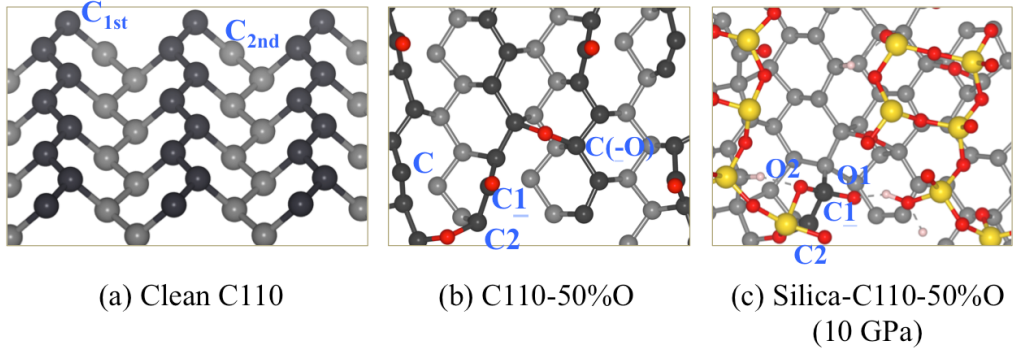

(a) Clean C110  (b) C110-50%O  (c) Silica-C110-50%O (10 GPa)

Figure 7. Top (darker balls) and second top layers of the clean C(110) surface (a), C110-50%O (b), and silica-C110-50%O optimized under 10 GPa (c). This figure shows the atom labels used for the BOP results in **Table 1**.

Table 2. Bond overlap population of C−C and Si−O bonds. The $C_{1st}$, $C_{2nd}$ are carbon atoms of the 1$^{st}$ and 2$^{nd}$ layers of the C(110) surface as shown in **Figure 7a**. C is a carbon atom on the top layer of the C(110) surface, C(-O) is the carbon terminated by oxygen on the C110-50%O surface shown in **Figure 7b**. C1 and C2 are the two carbon atoms shown in **Figure 7c**. The BOP is calculated as the average over atoms of the same kind in each considered system.

| Bond/system | Clean C(110) | C110-50%O | C110-50%O (10 GPa) |
| --- | --- | --- | --- |
| $C_{1st}$−$C_{1st}$ | 0.53 | 0.45 | 0.43 |
| $C_{1st}$−$C_{2nd}$ | 0.41 | 0.37 | 0.37 |
| C−C |  | 0.46 | 0.45 |
| C−C(−O) |  | 0.39 | 0.39 |
| C1−C2 | 0.53 | 0.36 | 0.32 |
| Si−O |  |  | 0.27 |

In order to understand the nature of the bonding in the silica-diamond system, the bond overlap population (BOP) was estimated to compare the bond strength of the C-C bonds in the clean C(110), 50%O coverage C(110), and silica-C110-50%O at 10 GPa. The BOP measures the electron density shared between two atoms by the Lobster code.[46] The results reported in **Table 2** show that the $C_{1st}$-$C_{1st}$ bonds (the darker atoms in **Figure 7**) are stronger than the $C_{1st}$-$C_{2nd}$ bonds. This is consistent with what has been reported by Peguiron *et al.* indicating that the C-C bond dissociation always occurs at the C-C bonds connecting the carbon atoms of the zigzag chain and those of the underlying layer.[31] When the surface is oxidized, the BOP values of the $C_{1st}$−$C_{1st}$ and $C_{1st}$−$C_{2nd}$ bonds are reduced from 0.41 e to 0.37 e, indicating that the C-C become weakened under oxygenation.[18] However, the value is still much higher than that of Si-O bonds (0.27e). Therefore, the Si-O bond breaking always occurs first during sliding. When C is bonded to two O atoms (C1), the BOP of the C1-C2 bond drops significantly from 0.53$e$ on the clean surface to 0.36$e$ on the oxidized C(110) and to 0.32$e$ in the silica-C(110) at 10 GPa. As a result, the bond breaking was observed during the AIMD simulation at 10 GPa (**Figure 6b**). Therefore, it is necessary to have C atoms bonded to two oxygen atoms to weaken the C-C bonds, as in the C1-C2 case. This is consistent with our recent DFT calculation which suggested that wear can appear on the C(110) surface at the C=O double bond or C-O-Si bidentate structure.[47]

Snapshots of the C001-50%O system during sliding in the harsh conditions are reported in **Figure S4**. It is clearly indicative that although Si-O-C bond formation occurs across the C(001)-silica interface, the lateral displacement leads to the dissociation of Si-O/C-O bonds, and the C(001) surface remains unaffected. This is consistent with the DFT calculation showing C(001) surface is less vulnerable to wear than C(110) one.[47]

In summary, the C-C bond dissociation on the diamond surfaces occurs in subsequent steps: (1) the O-H bond of the silanol group is dissociated, leaving a Si-O dangling bond, (2) the Si-O dangling bond is attached to the diamond surface, forming Si-O-C bridge, (3) when the C atom forms covalent bonds with two O atoms, it becomes weakened and the C-C bond breaking occurs.

## 3.4 Evolution of atomic wear in the presence of passivating species

Small molecules such as $O_2$, $H_2$, or $H_2O$ abundant in the working environment can diffuse into the sliding contact. When a C-C bond is broken, these small molecules could passivate the newly formed dangling bond. To investigate this possibility, we performed DFT calculations on the adsorption and dissociation of $O_2$, $H_2$, and $H_2O$ at the C-C breaking site on C110-50%O. The results indicate that the dissociation is thermodynamically favorable with reaction energies of -2.37, -3.08, and -2.14 for $O_2$, $H_2$, and $H_2O$, respectively (**Figure S5**). This result suggests that the dangling carbon created by the C-C bond breaking can be saturated by O, H and OH fragments of the dissociated molecules. Selected configurations explored during the sliding of O-added, H-added, and OH-added silica-C110 systems under 10 GPa load are reported in **Figure 8**.

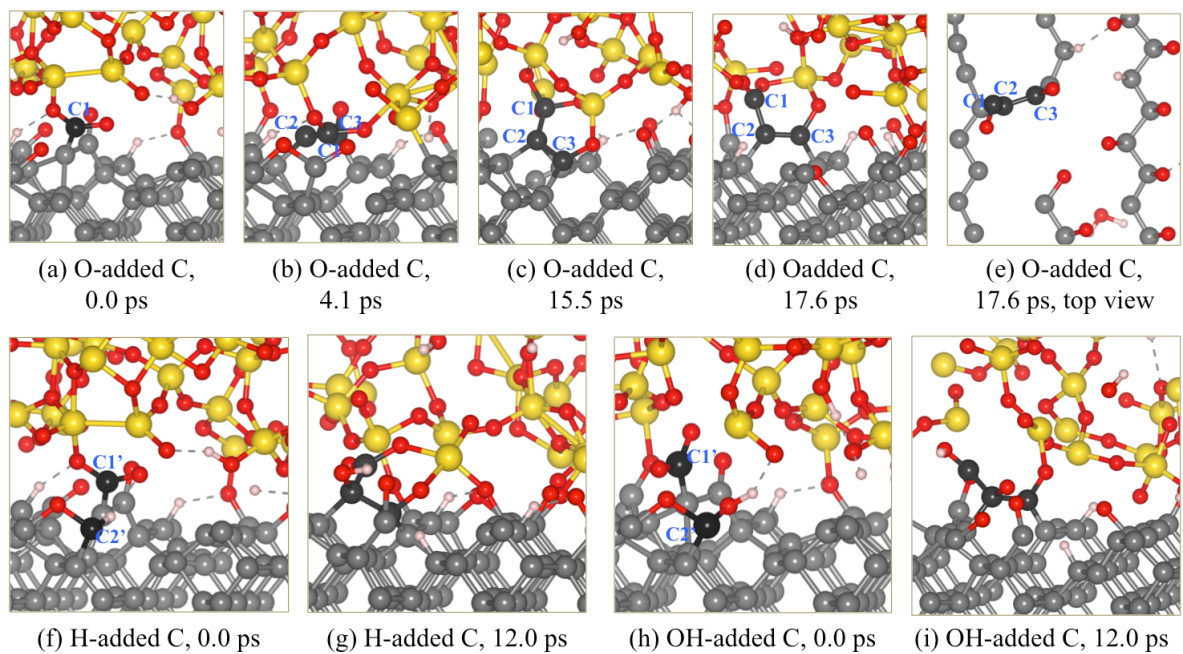

(a) O-added C, 0.0 ps  (b) O-added C, 4.1 ps  (c) O-added C, 15.5 ps  (d) O added C, 17.6 ps  (e) O-added C, 17.6 ps, top view

(f) H-added C, 0.0 ps  (g) H-added C, 12.0 ps  (h) OH-added C, 0.0 ps  (i) OH-added C, 12.0 ps

Figure 8. Formation of wear at the diamond-silica interface when O, H, and OH fragments terminate the C dangling bond (**Movie 7**).

When the carbon atom (C2) of the dangling bond is terminated, the bond recombination occurring in Figure 6d is not observed, and the sliding under 10 GPa leads to the dissociation of additional C-C bonds. Following the C1-C2 bond breaking, a second and a third C-C bond breaking event are promoted by the formation of Si-O-C bonds, respectively at 4.1 and 17.6 ps. The C-C bond dissociation results in the breaking of the zig-zag chain, as showed in **Figure 8e**. This can be a precursor of the wear formed through the displacement of carbon chains.[47] Similar C-C bond breaking is observed when the detached C is terminated by H and OH groups (**Figure 8f-i**), suggesting that environmental species play an important role in facilitating wear. The adsorption of the atmospheric molecules not only weakens the C-C bonds on the diamond surface but also terminates the dangling sites, thus preventing the C-C bond recombination and promoting wear.

## 4. Conclusions

The friction behavior and the atomistic wear mechanism of silica sliding against oxygenated diamond surfaces have been studied by AIMD simulations accompanied by atomic and electronic structure analyses. The obtained results can be summarized as follows:

1. The full coverage of diamond with oxygen is highly effective to reduce adhesion and the formation of chemical bonds across the silica-diamond interfaces. The outcome holds true for all three C(110), C(001), and $R$-C(111) surfaces at both 1 GPa and 10 GPa loads. The resistant forces in the case of the full O-coverage are even lower than those of the full H-coverage. This is due to the larger steric hindrance of oxygen and its electrostatic repulsion with the silica surface. The situation drastically changes for the lower O-coverage of different surface orientations. In particular, at 50% coverage, we observe the formation of Si-O-C bonds across the interface for the C(001) surface.

2. Under the harsh working conditions, chemical bonds are established at the interfaces of silica and half-passivated diamond surfaces except for the case of the *R*-C(111)-50%. However, the formation of Si-O-C bonds is not enough to induce the C-C bond breaking, which occurs only in the C(110)-50%O system when the C-C bond is weakened by the chemical bonds forming by two oxygen atoms at the same C site.

3. The recombination of the C-C broken bond can be prevented by the dissociative adsorption of passivating molecules present in the environment. The wear mechanism is then dominated by the detachment of carbon chains.

Our simulations indicate that full oxygenation is an effective technique for friction reduction and reveal the mechanical-chemical conditions to explain the wear formation in diamond-silica systems at partial O-coverage, which helps answer the question of how silica can polish diamond.

Acknowledgement

These results are part of the "Advancing Solid Interface and Lubricants by First Principles Material Design (SLIDE)" project that has received funding from the European Research Council (ERC) under the European Union's Horizon 2020 research and innovation program (Grant agreement No. 865633).

Supporting Information

Top and side views of oxidized diamond surfaces, Relaxed structures of silica-diamond systems at 1 and 10 GPa, Structures of molecular and dissociative adsorption of $O_2$/$H_2$/$H_2O$ on C(110)- 50%O. Movie 1-6 show the sliding of silica against 6 oxidized diamond surfaces at 1 GPa, Movie 7 shows the sliding of silica against C(110)-50%O at 10 GPa.

TOC Graphic

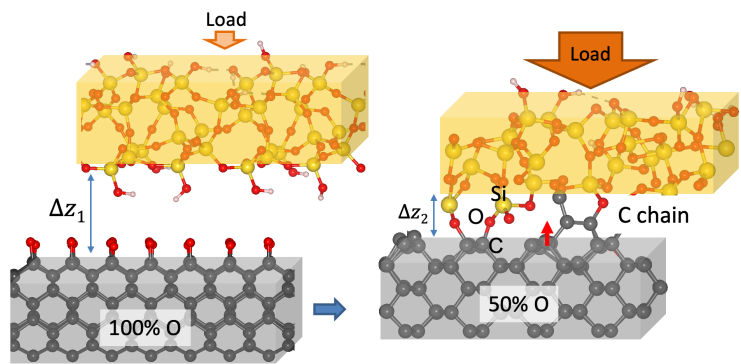